\pdfoutput=1 % only if pdf/png/jpg images are used
\documentclass[11pt,a4paper]{article}
\usepackage{graphicx, amssymb}
\usepackage{authblk}
\newcommand{\etaeff}{\eta_{\mathrm{eff}}}
\newcommand{\mm}{\mathrm{mm}}
\newcommand{\cm}{\mathrm{cm}}
\newcommand{\kV}{\mathrm{kV}}

\newcommand{\ns}{\mathrm{ns}}
\newcommand{\ps}{\mathrm{ps}}
\newcommand{\M}{\mathrm{M}}
\newcommand{\mum}{\mathrm{\mu m}}
\newcommand{\gasgem}{\mathrm{Ar/CO_{2}/CF_{4} = 45/15/40}}

\providecommand*\etal{\emph{et al.}}

\begin{document}

\title{\bf A novel fast timing micropattern gaseous detector: FTM}
\author[a]{R. De Oliveira}
\author[b]{M. Maggi\thanks{Corresponding author (Email: marcello.maggi@ba.infn.it)}}
\author[a]{A. Sharma}
\affil[a]{CERN, European Organization for Nuclear Research, Geneva, Switzerland}
\affil[b]{INFN Sezione di Bari, Bari, Italy}
\date{11$^{st}$ March 2015}

\maketitle

\begin{picture}(0.001,0.001)
\put(345,225){\makebox(0,0)[rb]{\normalsize CERN-OPEN-2015-002}}
\put(312,210){\makebox(0,0)[rb]{\normalsize INFN-15-01/BA}}
\end{picture}
\begin{abstract}
In recent years important progress in micropattern gaseous detectors has been achieved 
in the use of resistive material to build compact spark-protected devices. The novel idea presented 
here consists of the polarisation of WELL structures using only resistive coating. This allows a new 
device to be built with an architecture based on a stack of several coupled layers where drift  
and WELL multiplication stages alternate in the structure. The signals from each multiplication stage 
can be read out from any external readout boards through the capacitive couplings. Each layer provides 
a signal with a gain of $10^{4}\,-\,10^{5}$. The main advantage of this new device is 
the dramatic improvement of the timing provided by the competition of the ionisation 
processes in the different drift regions, which can be exploited for fast timing at the high luminosity 
accelerators (e.g. HL-LHC upgrade) as well as far applications like medical imaging.
\end{abstract}

\pagebreak[4]

\section{Introduction}\label{sec:Introduction}

Recent years have seen much progress on micropattern gaseous detectors (MPGD)~\cite{SauliArchana,Archana}, 
mostly under the RD51 coordination~\cite{rd51}.
These detectors have been introduced using well established photolithographic
technology on PCB supports.
However, they might suffer discharges due to inherent design feature: the very small distance between anode
and cathode electrodes limiting the gain of a single structure to $\sim 10^3$~\cite{Review}.
In Micromegas~\cite{uomega}
the problem of the spark occurrence between the metallic mesh and the readout PCB has been solved with the
introduction of a resistive layer deposition on top of the readout~\cite{resuomega}.  Recently, resistive
layers have been adopted in WELLs~\cite{Bencivenni}. Instead in GEMs~\cite{GEM} the adopted solution is to share
the gain among different multiplication stages~\cite{TripleGEM1,TripleGEM2} without the need of a resistive layer.

This class of detectors is now being exploited in many applications since: they exhibit good spatial and time resolution,
high rate capability~\cite{HighRate}; they are cost-effective and can be used for large sensitive area~\cite{CMSGEM};
they are flexible and have been used for different  geometries~\cite{KLOEGEM}. Though time resolution of a few nanoseconds
is perfectly adequate in several applications, it would represent a limiting factor in others, where more precise timing
is required, for example triggering at high luminosity and medical imaging. To improve the time resolution, the novel detector, described
in this paper and named Fast Timing Micropattern (FTM) detector, is based on a series of fully resistive WELL.
% CAT, or $\mu$CAT structures~[?].                                                                                                             
The use of only resistive layers to polarise the drift and the multiplication volumes allows the construction of consecutive
drift-multiplication volumes. In fact, the extraction of signals in any multiplication stage is possible since the overall structure
is transparent to the signal, which can be picked up by external electrodes located on top and bottom of the new device.
The timing of the ionisation processes in the respective drift volumes will then compete leading to a decrease
of the arrival time to any multiplication volume and consequently, a decrease of the
fluctuations and consequently an improvement of the time resolution.

In this paper we describe the working principle of this new detector, the FTM, of which a first prototype has been realised
in the TE-MPE-EM Workshop at CERN and operated successfully for the first time in December 2014. This is one of the implementations
described in the European Patent Application number 14200153.6.

\section{The working principle}\label{sec:principles}

\begin{figure}[hbt]
     \includegraphics[width=0.95\textwidth]{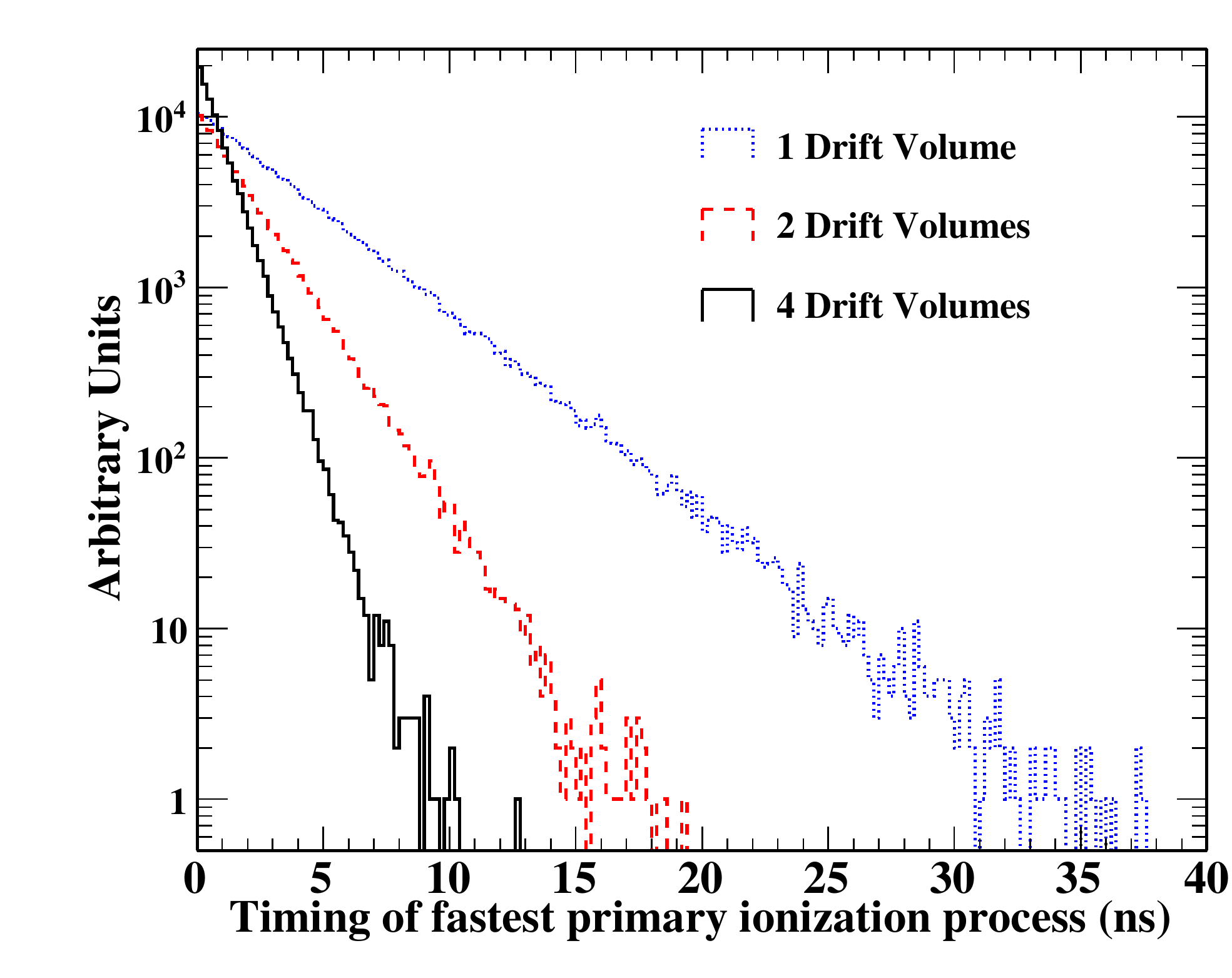}
\caption{Timing distribution of the fastest ionisation process.
The dotted line histogram represents the distribution for a single drift volume.
The dashed line histogram is the result obtained in a double layer configuration.
Finally the full histogram is the the distribution obtained in a configuration with four layers.}
\label{fig:exps}
\end{figure}

Micropattern detectors are characterised by a clear division of the various phases of the detection process, in
particular the primary ionisation creation and drift is separated by the multiplication phase and so the gain of the signal.
Thus, the time resolution is dominated by fluctuations of the nearest distance of the primary ionisation processes
to the region where the gain is acquired, $d_{near}$. Defining  $\lambda$ as the average number of primary clusters
generated by an ionising particle inside the gas, this distance follows a classical exponential distribution
$d_{near} = exp(-\lambda x)/\lambda$.  The drift velocity of the gas $v_{d}$ determines the arrival time,
the exponential behaviour is shown in Fig.~\ref{fig:exps} and contribution of $v_{d}$ to the time resolution is
\begin{equation}
        \sigma_{t} =  (\lambda v_{d} )^{-1}\ .
\label{eq:basic}
\end{equation}
Both gas parameters depend mainly on the gas mixture used in the device and, in addition, $v_{d}$ is also a
function of the electric field. Typical values for gases employed in MPGDs are $\lambda = 3~\mm^{-1}$ and $v_{d}$
up to  $0.1~\mm/\ns$ leading to few $\ns$ time resolution with the best choice of gas mixtures and operating voltages.
By comparison, the contribution of the gain fluctuation is governed by the gas time constant
$(\etaeff v_{d})^{-1}$, where $\etaeff$ is the effective Townsend coefficient. In the multiplication volume the field
is very high ($\sim 50\,-\,100~\kV/\cm$) and consequently  $\etaeff > 100~\mm^{-1}$ and $v_{d} > 0.25~\mm/\ns$ with a resulting
contribution to the time resolution much below $100~\ps$.

\begin{figure}
\includegraphics[width=0.7\textwidth, viewport=0 0 500 400]{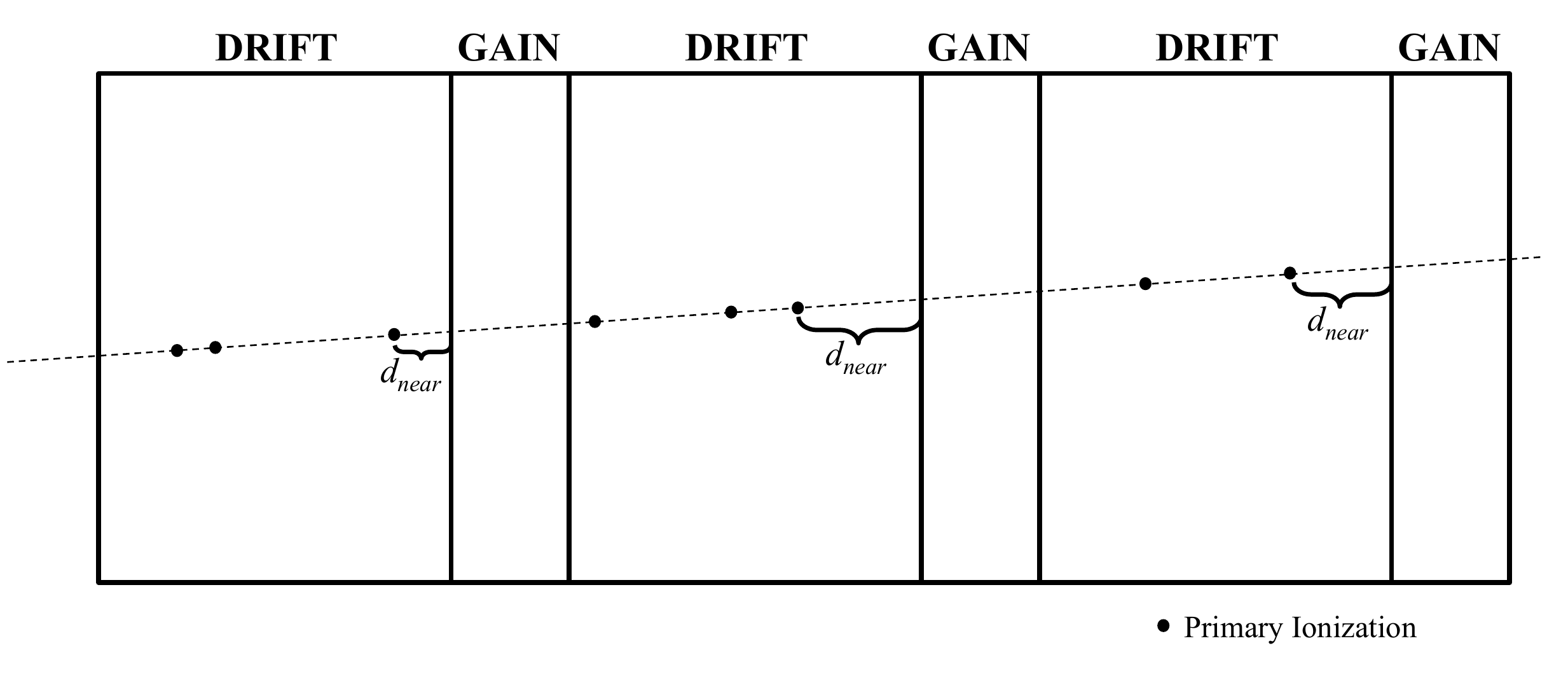}
\caption{Schematic of the working principle of the FTM.
Drift and gain processes alternate in the overall configuration that is a stack of several detection layers.
Electrons from the primary ionization clouds drift towards the multiplication volumes and timing is determined
by the cloud nearest to the respective gain region, which is represented by the minimum of the layer's $d_{near}$.}
\label{fig:prschema}
\end{figure}

\begin{figure}[hbt]
     \includegraphics[width=0.95\textwidth]{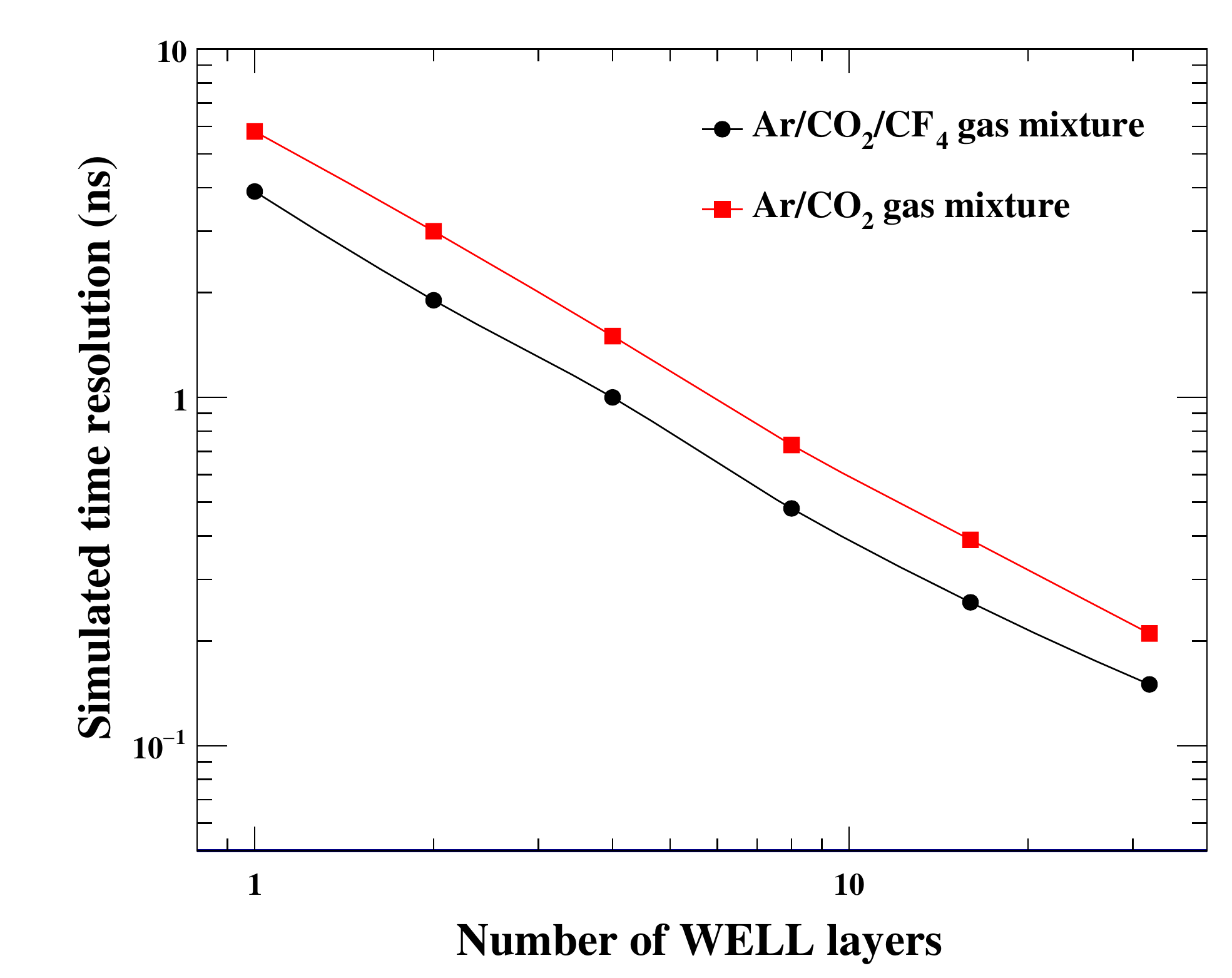}
\caption{Simulated time resolution of the FTM device as a function of the number of WELL layers.
The simulation has been performed with two standard gas mixtures: the full squares represents the time
resolutions with a mixture of Ar/CO$_{2}$ = 70/30; the full circles those obtained with $\gasgem$.}
\label{fig:nwell}
\end{figure}

In order to improve the time resolution a new configuration is proposed and is shown in Fig.~\ref{fig:prschema}. The
improvement is obtained by using several drift regions each one coupled to its multiplication stage, which is realised with a
fully resistive WELL structure. In Fig.~\ref{fig:exps} results of a simulation are shown for configurations using up to four drift regions,
which already demonstrate the benefit. The reduction of the time resolution, in fact, is proportional to the number of the
layers $N_{D}$ employed, leading Eq.~\ref{eq:basic} to be modified as follows:
\begin{equation}
        \sigma_{t} =  (\lambda v_{d} N_{D})^{-1} \ ;
\label{eq:Nbasic}
\end{equation}
The behaviour is also evident in Fig.~\ref{fig:nwell}, where the results of a simulation of timing response,
taking into account also the micro behaviour of the drift and avalanche formation with GARFIELD~\cite{GARFIELD}
is demonstrated.

\section{The detector layout and test set-up}\label{sec:detector}

\begin{figure}
\includegraphics[width=0.95\textwidth,viewport=0 0 340 400]{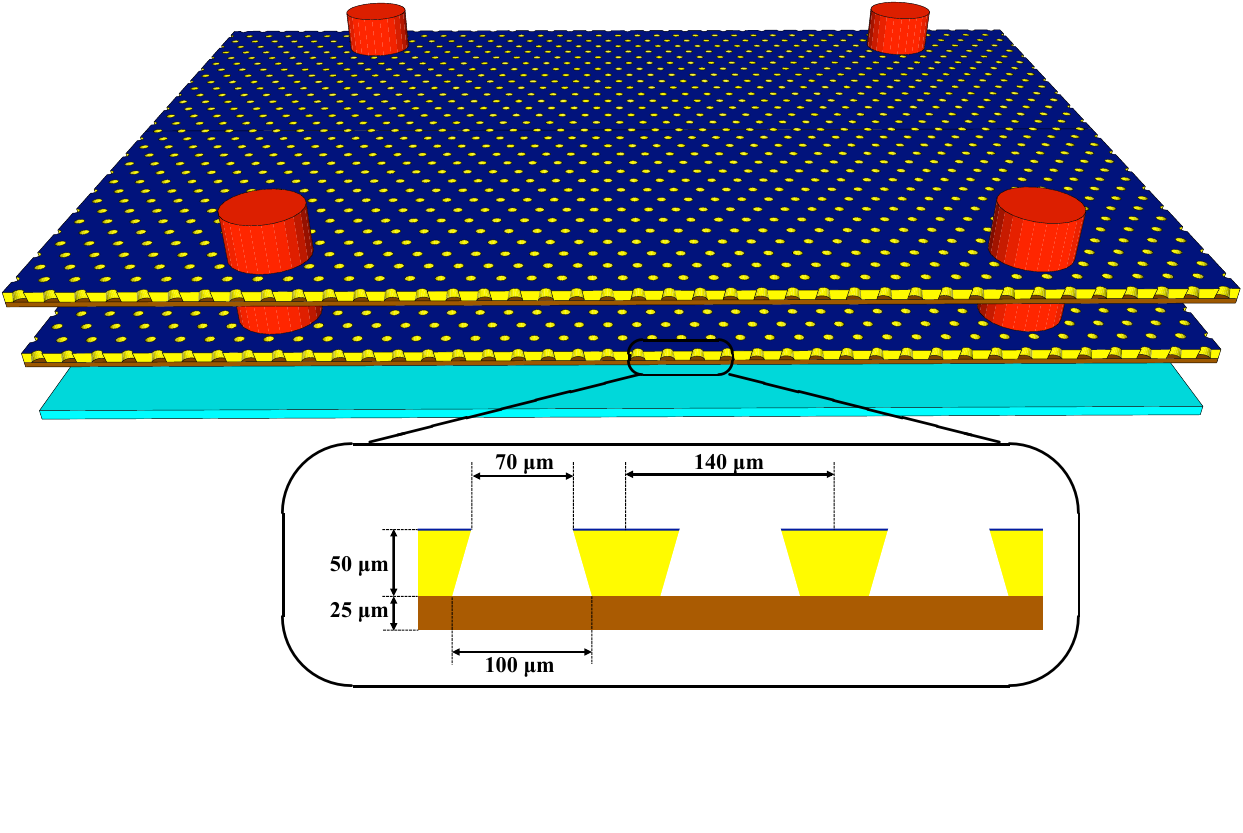}
\caption{Drawing of the first implementation of the FTM. The  basic structure used to build the prototype consists of two layers of full resis\
tive WELL with
DLC coating on the top (dark blue surface) on the perforated foils (yellow volumes) and amplification volumes are closed by the antistatic pol\
yimide foils
(brown volume). The red cylinders are the pillars. In light blue the pick-up electrode is represented. In the zoomed area is visible the detai\
l in 2D of the fully resistive WELL.}
\label{fig:sketch}
\end{figure}

The detector layout of the prototype is based on the concept of the micro-Resistive WELL
detector~\cite{Bencivenni}:  two layers are used for this first implementation,
which is sketched in Fig.~\ref{fig:sketch}. The multiplication volume is based on
of a pair of polyimide foils stacked due to the electrostatic force induced by
the polarisation of the foils. The perforated foils are a $50~\mum$ thick Apical KANECA,
which is initially coated with diamond-like carbon (DLC)  techniques. This coating technique
is used to provide, to the perforated polyimide foil, a potential gradient
with resistive layers of the order of $800~\M\Omega/\square$, replacing the copper coating.
The holes are inverted truncated cones with the two bases of $100~\mum$ and $70~\mum$ diameter.
The pitch between holes is of $140~\mum$. The second foils are the $25~\mum$ thick XC Dupont
Kapton, which have a resistivity of the order of $2~\M\Omega/\square$. The drift volumes are
$250~\mum$ thick. The planarity is ensured by a set of pillars obtained by PCB technique with
photoimageable coverlay. The pillars' diameter is $400~\mum$ with a pitch of $\sim 3.3~\mm$.
The spacing around the active region is ensured by the use of two $125~\mum$-thick polyimide foils.

\begin{figure}[hbt]
     \includegraphics[width=0.95\textwidth,viewport=0 0 600 400]{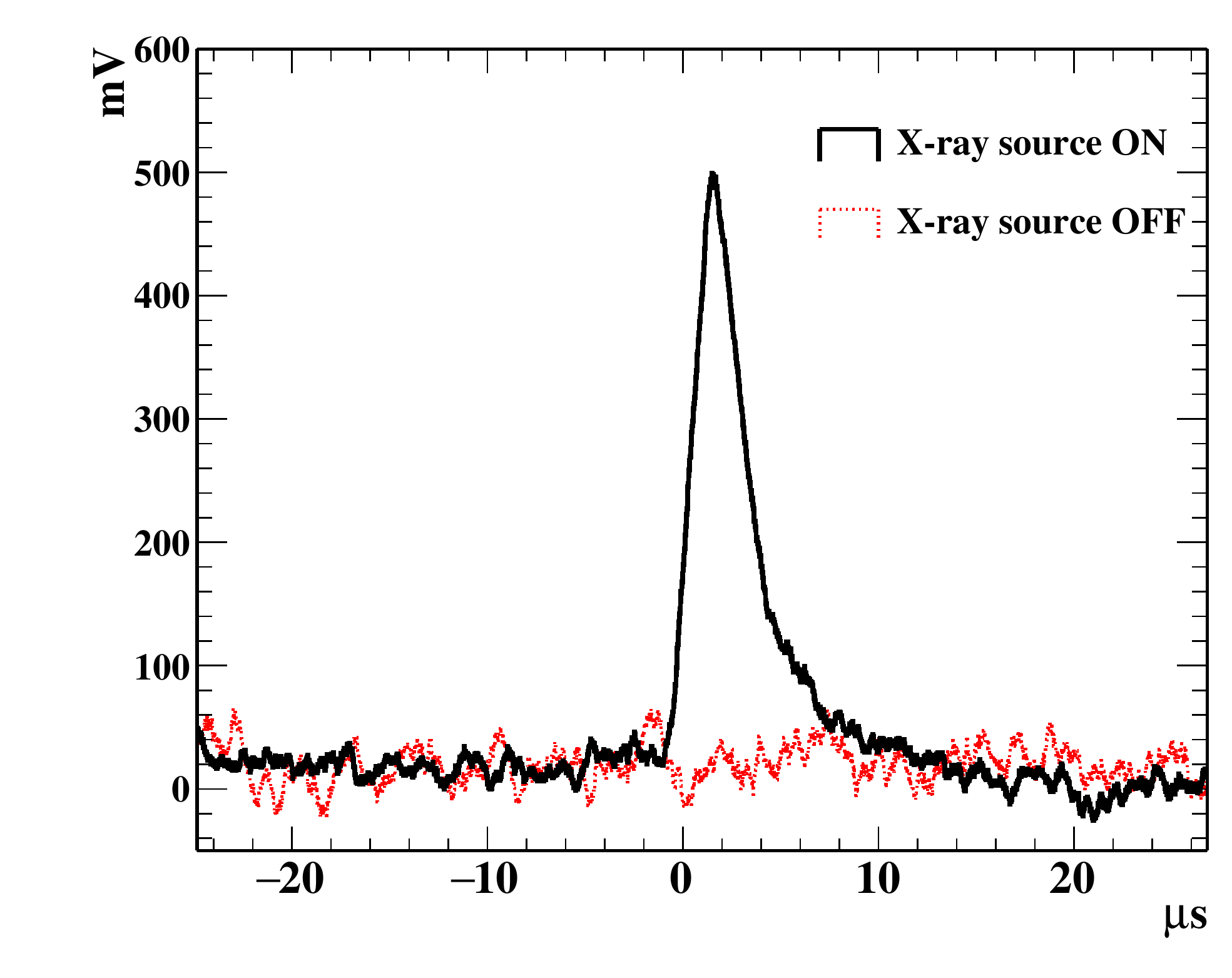}
\caption{Average FTM signal with X-ray source on (full line) and off (dotted) line. }
\label{fig:signal}
\end{figure}

The employment of only resistive layers in the architecture allows signals coming from the layers  to be externally extracted thanks 
to the resulting transparency of polarising electrodes. The prototype has been operated with a $\gasgem$ gas mixture. The different regions have been
polarised with the CAEN N1470 power supply in order to give an electric field of $2~\kV/\cm$ for the drift regions and $100~\kV/\cm$ for the 
amplification regions for a gain between $10^4$ and $10^5$. The signals from the readout electrode
have been sent to the low-noise charge-sensitive ORTEC PC142 preamplifier.
The resulting inverted signal outputs have been amplified by the ORTEC 474 NIM module and
acquired with a Tektronix TDS 2024C oscilloscope. An X-ray source has been used in order
to demonstrated the capability of signal extraction, and the results are shown in Fig.~\ref{fig:signal} where an average signal is shown
in presence and absence of the source. This is the first signal coming from a FTM device allowing the implementation of the
principles described in this paper. The detailed studies of the performance for this class of devices has started and will be the subject of future
papers.

\section{Conclusions and outlook}
In this paper we have described a novel class of micropattern gaseous detectors for fast timing applications called the FTM.
The construction feasibility has been demonstrated by building a first working prototype. We expect that this technique can
be exploited for applications in high energy physics experiments, particularly for upgrades at LHC where sub nanosecond time
resolutions are critical for particle identification and vertex separation. Other applications include X-ray diffraction
studies and fast time-resolved measurements offer excellent medical imaging opportunities. In combination with an X-ray
convertor and FTM and a visible photocathode shows great promise for use
in digital mammography. Other applications include X-ray astronomy by exploiting time resolution of the FTM and selective
sensitivity to soft X-rays.
For further studies, the subject of future publications, we are building prototypes with several layers, and in parallel
investigating larger size cost effective production techniques.

\section*{Acknowledgments}
We are grateful to Silvia Franchino for assembling the device and assisting with the measurements.

\pagebreak[4]
\pagebreak[4]
\pagebreak[4]
\pagebreak[4]
\pagebreak[4]
\end{document}